# A matching-based heuristic algorithm for school bus routing problems

Ali Shafahi[1], Zhongxiang Wang[2*], Ali Haghani[3]

[1]Department of Computer Science, University of Maryland - College Park, MD 20742, Email: ashafahi@umd.edu

[2*]Department of Civil and Environmental Engineering, University of Maryland - College Park, MD 20742, Email: zxwang25@umd.edu (Corresponding author)

[3]Department of Civil and Environmental Engineering, University of Maryland - College Park, MD 20742. Phone: (301) 405-1963, Fax: (301) 405-2585. Email: haghani@umd.edu

**Abstract**
School bus planning problem (SBPP) has drawn much research attention due to the huge costs of school transportation. In the literature, the SBPP is usually decomposed into the routing and scheduling subproblems due to its complexity. Because of the nature of the decomposition, even if all the subproblems are solved to optimality, the final solution may not be as good as the solution from the integrated model. In this paper, we present a new approach that incorporates the scheduling information (namely the trip compatibility) into the routing stage such that the interrelationship between the subproblems is still considered even in the decomposed problems. A novel two-step heuristic adopting the trip compatibility idea is presented to solve the school bus routing problem. The first step finds an initial solution using an iterative minimum cost matching-based insertion heuristic. Then, the initial trips are improved using a Simulated Annealing and Tabu Search hybrid method. Experiments were conducted on randomly generated problems and benchmark problems in the literature. The result shows that our two-step heuristic improves existing solutions up to 25% on the benchmark problems.

# 1. Introduction

According to the National Center for Education Statistics, the *U.S. spent over $23 billion on public student transportation in 2012-2013* this is about $914 per student (National Center for Education Statistics, 2016). Due to the vast amount of funds being invested in school transportation, improving operational efficiency even a little could result in huge savings. The primary objective of designing a school bus transportation system, from the operator's perspective, is to safely transport all students with minimum cost while satisfying constraints such as maximum ride time, vehicle capacity, and time window. This problem is formulated as a school bus planning problem (SBPP). Due to this importance, many studies have focused on improving the efficiency of school transportation plans.

A school transportation plan consists of the routing plan and the scheduling plan. The routing plan is the joint of a set of *trips* where each trip starts from a school and sequentially visits a set of stops that are exclusive to that school. The scheduling plan is made of a set of *routes* where each route starts from the depot and sequentially serves a set of compatible trips. An ordered trip pair is *compatible* if a bus after finishing the preceding trip, has enough deadhead (travel time from the last stop in the preceding trip to the first stop of the successive trip) to reach the first stop of the successive trip before its service start time. The cost of a school transportation plan mainly includes the acquisition cost of a bus, bus driver employment cost, and operations cost (like gas, maintenance, etc.). The first two costs are fully determined by the number of buses used, and the last one highly depends on the total travel time/distance. Therefore, the objective of the SBPP is to minimize the weighted sum of the number of buses and the total travel time/distance. The former is of higher priority than the latter because the acquisition cost of a bus and bus driver employment can easily overwhelm the marginal cost of the travel distance increase.

In this study, we develop a two-step heuristic algorithm for solving the school bus routing problem with the consideration of trip compatibility. The first step builds an initial set of trips using an insertion-based heuristic that adds stops to trips by iteratively solving minimum cost matching problems. The second step improves the initial trips by performing chain exchange guided by Simulated Annealing and Tabu Search. This fast heuristic algorithm can generate a set of trips to serve all stops for every single school. Given the trips for all schools, the multi-school bus scheduling can be solved as an assignment problem under the condition that the trip start times are known (Kim et al., 2012). Following this structure, the whole SBPP can be solved. Our solution approach is general and can be used for solving other variants of vehicle routing problems that include routing and scheduling such as dial-a-ride and pickup and drop-off and delivery problem with time window. We show the effectiveness of our proposed methodology on benchmark problems from the literature. The rest of the paper is structured as follows: In section 2, we review some of the literature regarding solution methods for solving school bus routing problem and related variants of vehicle routing problems along with the research gap and our contribution. In section 3, we present our solution algorithm. Section 4 analyzes the results and illustrates the performance of our proposed solution algorithm on randomly generated problems and benchmark problems. Finally, section 5 concludes the paper.

# 2. Literature Review

As pointed by (Park and Kim, 2010), the majority of the school bus planning research decompose the SBPP into subproblems and only focus on one of the subproblems. While a few recent studies such as (Park et al., 2012) and (Bögl et al., 2014) formulated integrated mathematical models for the SBPP. However, due to the complexity of the models, they decomposed those models into sub-problems during the solution process. A conceptual flowchart of routing and scheduling is shown in Figure 1 where bus stop selection and school bell times are assumed to be predetermined.

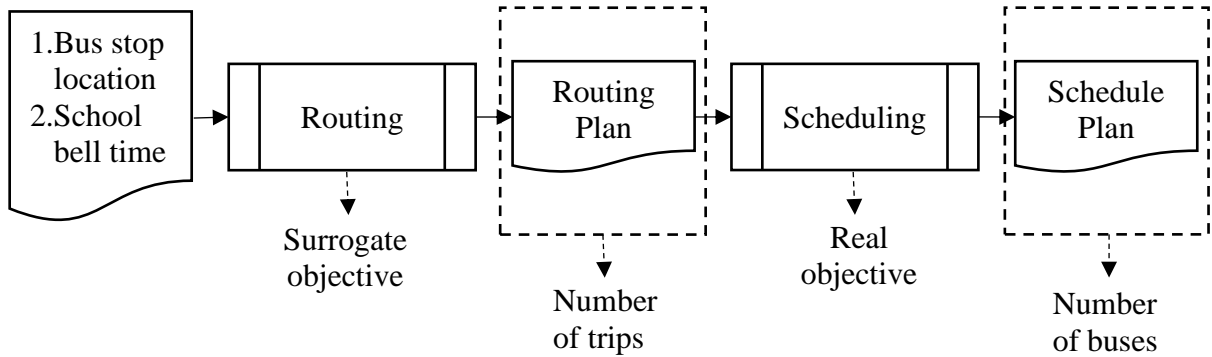

Figure 1 Flowchart of routing and scheduling

As discussed above, the primary objective of the SBPP is to minimize the number of buses used. However, due to the precedence of the routing step, this objective cannot be accounted for in the process of routing. As a consequence, a surrogate objective is required to solve the routing problem. Different surrogate objectives will lead to different routing plans, which will further result in different schedule plans. Traditionally, the surrogate objective for routing is to: 1) minimize the number of trips; or 2) minimize the total travel distance or the travel time (Schittekat et al., 2013; Faraj et al., 2014; Kinable et al., 2014); or 3) a mixture of 1 and 2 (Díaz-Parra et al., 2012; Caceres et al., 2017). Those surrogate objectives also serve as the evaluation criteria in many heuristics like the insertion cost estimation in trip construction algorithms and the exchange move goodness estimation in local search algorithms.

Multi-school routing problems are often broken down into a group of single-school routing problems due to the problem size. For solving single-school routing problems, both exact and heuristic algorithms have been proposed and implemented. The exact approaches use mathematical modeling and commercial MIP solvers to solve the problem. They are applied to cases where the single school problem is small and has a limited number of students and stops. The modified assignment problem is one of the first exact approaches. Bektaş and Elmastaş (2007) duplicated the depots as dummy nodes and transformed the school bus routing problem into a graph problem of finding node-disjoint paths. The problem they solved had only 29 stops. Most studies, however, focus on solving larger size problems. Due to the complexity of the routing problem, these studies develop heuristic algorithms for solving the single school routing problem. A popular heuristic approach is to formulate the school bus routing problem as a set partitioning problem and solve it by a restricted column generation approach. This method was adopted by Braca et al. (1997), Riera-Ledesma and Salazar-González (2013) and Kinable et al. (2014). Some other widely used heuristics for solving the school routing problem or the vehicle routing problem in general are the Clarke and Wright saving method, insertion method (Christofides et al., 1979; Bramel and Simchi-levi, 1992; Braca et al., 1997; Corberán et al., 2002), sweep algorithm (Park et al., 2012), Simulated Annealing (SA) (Spada et al., 2015), Deterministic Annealing (Braekers et al., 2014), Tabu Search (TS) (Nanryand Barnes, 2000; Cordeau and Laporte, 2003; Fu et al., 2005; Spada et al., 2015; Pacheco and Martí, 2006; Paquette et al., 2013), Genetic Algorithm (GA) (Thangiah and Nygard, 1992; Prins, 2004; Díaz-Parra et al., 2012; Kang et al., 2015), Ant Colony (Yao et al., 2016) and GRASP (Schittekat et al., 2013; Faraj et al., 2014). GRASP (Greedy

Randomized Adaptive Search Procedure) can be considered as an improved version of the Clarke and Wright savings method. When building the greedy saving list, GRASP adds randomness to prevent the heuristic from getting stuck in a local optimum point. Schittekat et al. (2013) and Faraj et al. (2014) both adopted this algorithm.

The basic idea of an insertion-based routing algorithm is to add (insert) stops to trips based on some insertion costs. The routing plans will be completed when all the stops are inserted into trips, and no stops remain. Insertion of stops to trips can be done either sequentially or in parallel. The sequential procedure builds one trip at a time while the parallel procedure builds multiple trips together. One example of the sequential procedure is the Location Based Heuristic proposed by Bramel and Simchi-levi (1992). Christofides et al. (1979) proposed a two-phase algorithm to apply sequential procedure followed by a parallel procedure. A similar stop-by-stop routing algorithm is the sweep algorithm, which sequentially inserts a stop based on its polar angle. It was adopted by Park et al. (2012) where at each stop insertion, a neighbor traveling salesman problem was solved, and 2-opt improvement procedure was applied to construct a good trip. They also developed benchmark problems which we use in this study. There are many other heuristic algorithms that apply local search post-improvement steps to improve the trips further. Sometimes these post improvement steps borrow ideas from insertion-based routing and modified swap algorithms. One example is from Corberán et al. (2002), which combined insertions and swaps to improve trips that were initially generated using an insertion-based heuristic.

Some other solution algorithms perform the post-improvement step using well-known meta-heuristics such as Tabu Search, Simulated Annealing, Deterministic Annealing and Genetic Algorithm. All of them can guide the direction of local search and avoid getting stuck in local optima. Nanryand Barnes (2000), Cordeau and Laporte (2003), Fu et al. (2005), Pacheco and Martí (2006), and Paquette et al. (2013) applied the Tabu Search to solve school bus routing problem or other related vehicle routing problem like dial-a-ride problem or pickup and dropoff problem. Spada et al. (2005) adopted Simulated Annealing to maximize the level of service given a fixed number of available buses. Braekers et al. (2014) used Deterministic Annealing to solve a multi-depot heterogeneous dial-a-ride problem. Deterministic Annealing is a variant of Simulated Annealing, which accepts a neighbor solution if that solution only worsens the objective value within a threshold and this threshold decreases over time. The most common local search mechanisms include edge exchange and chain exchange. Edge exchange removes some edges from a trip and finds a better reconnection from other trips or sometimes itself. A famous example is a 2-opt algorithm, which can effectively eliminate crossing edges from a trip (Babin et al., 2007). Chain exchange moves vertices instead of edges. A famous chain exchange algorithm is Or-opt (Or 1976).

Another approach for solving the routing problem is based on decomposition. It is usually done using clustering methods. Based on the time that clustering is applied, these families of algorithms can be classified into 1) Route First, Cluster Second (RFCS) or 2) Cluster First, Route Second (CFRS) algorithms. In the RFCS algorithm, initially, one long trip is constructed by finding a Hamiltonian cycle. This step is usually referred to as finding a 1-TSP[1], and it is done by relaxing the capacity and maximum ride time constraints. In the next step, this trip is partitioned into several smaller trips which satisfy the capacity and maximum ride time constraints. One example of RFCS is Space Filling Curve with Optimal Partitioning from Bowerman et al. (1994). The Cluster First, Route Second (CFRS) algorithm first groups stops into different clusters while making sure the capacity constraints are not violated in each cluster. Then a 1-TSP is solved for each cluster. Some local search steps are added if maximum ride time constraints are violated. CFRS was adopted by Bowerman et al. (1995) to solve urban school bus routing problem with multiple objectives.

---

[1] 1-TSP: Traveling Salesman Problem with one trip

A new perspective of solving school bus routing problem was presented by Ellegood et al. (2015) using continuous approximation model. The model was tested in the Windsor School District, Missouri with 5 schools and 2301 bus riders. The result shows that the mixed-load plan can save 8.4% expected travel distance than the current operation.

Matching algorithms have proven to be suitable for solving large size problems because they could be solved optimally in polynomial time using methods such as the Hungarian method with $O(n^3)$ (Munkres, 1957). They have been successfully applied to school transportation planning problems (Forbes et al., 1991; Wark and Holt, 1994; Harvey et al., 2006; and Shafahi et al., 2017a).

One downside of the routing and scheduling decomposition strategy, which also exists in the algorithms that fall into this decomposition framework, is that the interrelationship between the subproblems is neglected. Due to the nature of decomposition, optimal trip plan of the routing stage, do not necessarily result in a good scheduling plan. To counter this deficiency, Shafahi et al. (2017b) and Wang et al. (2017) proposed to minimize the number of trips and total travel time while maximizing trip compatibility in the routing stage. Their strategy is shown to produce better routing plans and better scheduling plans. However, their method relied on solving a MIP problem using the branch and bound algorithms implemented in commercial solvers. They can only solve the single school problem with up to 30 stops. In reality, schools have much more stops. Since they did not have an efficient algorithm, their models were not able to solve real-world large-scale problems. In this paper, we develop an efficient heuristic algorithm to solve the school bus routing problem with the consideration of trip compatibility. This 'look ahead' strategy introduces the scheduling information (namely trip compatibility) into the routing stage. With the 'look ahead' information, our algorithm can construct better routing plan with more compatible trips, which will lead to a better scheduling plan using a fewer number of buses. The trip compatibility is considered in both the trip construction stage and the post-improvement stage.

## 3. Methodology

We develop a novel Minimum Cost Matching with Post Improvement (MCMPI) algorithm solve the school bus routing problem with the homogeneous fleet. We assume that students from different schools cannot be assigned to the same trip (i.e., single-load). Furthermore, the school bus depot location, the number of students at each stop, and the school bell times are assumed to be predetermined. A detailed description of this solution algorithm is provided in the next subsection. For illustration purposes, we present our algorithm as an algorithm for afternoon trips in which the bus service start time equals to the school dismissal time, which makes it easier to illustrate the trip compatibility. However, due to symmetries, constructing morning trips is similar to constructing afternoon trips since either of them can be obtained merely by reversing the other one. MCMPI is decomposed into two parts: the minimum cost matching (MCM) trip generation algorithm and the post-improvement (PI) simulated annealing based algorithm. The basic procedure of MCM and PI are described in the next two subsections.

3.1. Minimum Cost Matching Trip Generation Algorithm

The minimum cost matching (MCM) trip generation algorithm is an insertion-based routing algorithm where the stop insertion decisions are made by solving minimum cost matching problems. In a multi-school setting, we decompose the problem by school and solve each single-school routing problem one at a time. The inputs to each single school problem are the stops and students of the school and the school dismissal times for all schools. The dismissal times for other schools are used to calculate the compatibility of trips. Since we are solving the problem one school at a time, we can reduce the problem size by considering trip-school compatibilities as opposed to trip-trip compatibilities. If a trip is compatible with another school, then this trip and a trip from the other school can be served by one bus. This compatibility is beneficial regarding reducing the number of buses. It introduced the scheduling information into the routing stage without explicitly solving the scheduling problem. The MCM routing can be either done sequentially or in parallel. The MCM pseudo-code is shown in Algorithm 1 where all notations are summarized in Table 1. A graphical example of MCM in illustrated in Figure 3.

Table 1 Notations for minimum cost matching routing algorithm

| Variables | Description |
|---|---|
| $MQ_{i,j}$ | Remaining vehicle capacity of trip $j$ after inserting stop $i$ |
| $MC_{i,j}$ | Trip compatibility criterion of trip $j$ after inserting stop $i$ |
| $MT_{i,j}$ | The travel time of trip $j$ after inserting stop $i$ |
| $x_{s,t}$ | Binary variable, equals 1 if stop $t$ is directly succeeding stop $s$ one a trip; 0, otherwise, which is updated after the matching problems are solved. |
| $y_{i,j}$ | Binary variable, equals 1 if the insertion of stop $i$ to trip $j$ is selected; 0, otherwise. Due to the total unimodularity of the matching problem, $y_{i,j}$ can be relaxed to a non-negative continuous variable. |
| **Parameters** | **Description** |
| *Schools* | Set of Schools |
| $SB_k$ | School dismissal time for school $k$ |
| *Stops* | Set of stops for the school that we are solving the routing problem for |
| $Stops_j$ | A list of stops building trip $j$ |
| *Nsch* | Number of schools |
| *Nst* | Number of stops for the current school |
| *MRT* | Maximum ride time |
| *Cap* | Bus capacity |
| *N* | Set of trips |
| *n* | Number of trips ($|N|$) |
| *R* | Un-routed (un-assigned) stop set |
| *r* | Number of un-routed stops ($|R|$) |
| $IC_{i,j}$ | Insertion cost of inserting stop $i$ into trip $j$ |
| $\alpha_Q$ | Coefficient of the remaining (waste) vehicle capacity in insertion cost |
| $\alpha_C$ | Coefficient of trip compatibility in insertion cost |
| $\alpha_T$ | The coefficient of travel time in insertion cost |
| $O_j$ | The school that trip $j$ belongs to |
| $l_{i,j}$ | Last stop in trip j after inserting stop $i$ |
| $D_{s,t}$ | The travel time between node $s$ to $t$ without a pickup and drop off time |
| $PT_k$ | Student pickup time at school $k$ |
| $DT_i$ | Student drop-off time at stop $i$ |
| $q_i$ | Number of students at stop $i$ |

**Algorithm 1 MCM**
1. initialize the number of trips (n)
2. cluster stops into n clusters [by k-means algorithm]
3. find the 'furthest' stop to the school in each cluster
4. initialize the trip set (N) and the un-routed (unassigned) stop set (R)
5. **while** $R \neq \emptyset$ **do**
   5.1 construct the insertion cost matrix
   5.2 solve the minimum cost matching problem
   5.3 **if** (feasible insertion exists) **then**
        a. update N, R, n, r
   **else**
        b. initialize a new trip to N, update R, n, r

*3.1.1. Initialization*

     Initialization refers to steps 1 – 4 of the MCM algorithm. The MCM algorithm starts with a predefined number of trips. This predefined number should not be set to a high value because the final number of trips of MCM would never go below this value. We experiment with a different initial number of trips. The highest number still should be the lower bound on the actual number of trips. Therefore, we select the value to be equal to $\lceil \sum_{i \in Stops} q_i / Cap \rceil$. The smallest value for the initial number of trips would be 1. For higher values of the initial number of trips, several trips will be initialized simultaneously, making the MCM process a parallel routing process. We call this the Parallel Minimum Cost Matching (PMCM) routing algorithm. On the other extreme case, if the initial number of trips is set to one, only one trip will be initialized, and the algorithm becomes the Sequential Minimum Cost Matching (SMCM) routing. Both PMCM and SMCM initialize trip sets by having the school stop as the first stop and adding a stop that is far from the school location. We set the first stop (second if we are also counting the school stop) to be the furthest stop from the school. The intuition behind this move is that those stops which are furthest away are more influential to the maximum ride time than the closer ones. We want to have trips that do not violate the maximum ride time constraints. So, we should first take care of these stops. When we have more than one initialized trip (PMCM), to prevent assigning stops that are close together but far from the school to different trips, we use a Cluster First, Assign Second scheme. The clustering step makes sure that the 'furthest' stops used for initialization are more scattered. For clustering the stops, we use pair-wise distance and the k-means algorithm from scikit-learn (Pedregosa et al. 2011) where k equals the number of initialized trips (n). Once the clustering is done, we assign the furthest stop in each cluster to each trip.

*3.1.2. Trip construction and stop assignment*

     This subsection refers to step 5 of the MCM algorithm. Once the trips are initialized, the remaining task is to insert the remaining (un-routed) stops into trips. To achieve this goal, we iteratively solve matching problems until all stops are inserted. During each matching round, we either insert remaining stop(s) to trips or add a new trip. Which case happens, depends on the whether the insertions are feasible. To solve the matching problem, we construct a hypothetical bi-partite graph. One set of the graph vertices represent the trips, and the other set represents the remaining stops. The weight between vertices $i$ and $j$ represents the costs associated with adding (inserting) stop $i$ to trip $j$. An illustration of the bipartite graph is shown in Figure 2 with five remaining stops and two trips. There are in total 10 possible assign (insertion) pairs. After solving

the matching problem, two pairs are selected (inserting stop 2 into trip 1 and inserting stop 7 into trip 2) such that the total cost is the minimum. After updating the insertion pairs, we are left with three remaining stops and two trips. Since only one insertion is feasible (inserting stop 1 into trip 1), only this insertion will be selected and updated. The algorithm will continue with two remaining stops and two trips. The costs/weights needed for solving the matching can be stored in a matrix that we call the insertion cost matrix. Section 3.1.2.1 explains in more detail how the costs are calculated. Section 3.1.2.2 explains how to find the best insertion by solving a matching problem under the satisfaction of the feasibility constraint.

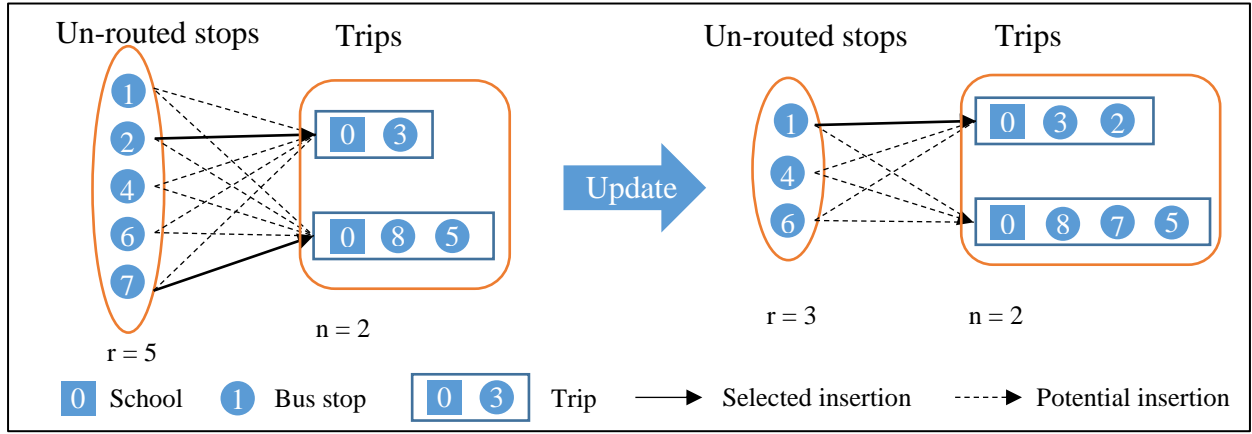

Figure 2 Illustration of the bipartite graph for matching problem

### 3.1.2.1 Constructing the insertion cost matrix.

A core task in minimum cost matching is constructing the $r \times n$ insertion cost matrix, which evaluates the cost of inserting a stop $i$ from un-routed stop set $R$ (of size $r$) into a trip $j$ in trip set $N$ (of size n). This construction is step 5-1 of the MCM algorithm. For PMCM, the insertion cost is a $r \times n$ matrix while in SMCM, it is a $r \times 1$ vector. The insertion cost (*IC*) depends on factors such as augmented travel time, the number of students (remaining bus capacity), and compatibility and they are defined in Equations (1)-(4).

$$IC_{i,j} = \alpha_Q MQ_{i,j} + \alpha_C MC_{i,j} + \alpha_T MT_{i,j} \;\; \forall i \in R, j \in N \quad (1)$$
$$MQ_{i,j} = Cap - \sum_{s \in Stops_j \cup i} q_s \;\; \forall i \in R, j \in N \quad (2)$$
$$MT_{i,j} = \sum_{s \in Stops_j \cup i} \sum_{s \in Stops_j \cup i | s \neq t} x_{s,t} D_{s,t} + DT_{O_j} + \sum_{s \in Stops_j} PT_s \;\; \forall i \in R, j \in N \quad (3)$$
$$MC_{i,j} = Nsch - \sum_{t \in Schools} 1\{SB_{O_j} + MT_{i,j} + D_{l_{i,j},t} \leq SB_t\} \;\; \forall i \in R, j \in N \quad (4)$$

The insertion cost between a stop and a trip is a weighted average of three factors, namely remaining bus capacity, trip compatibility to all other schools and the travel time. The remaining bus capacity defined in Equation (2) is the bus capacity minus the sum of students in the stops in trip $j$ and stop $i$. The intuition behind this factor is that we prefer to assign a stop to a trip if this assignment helps in filling up the trip (bus). The travel time is the sum of travel time between the nodes (stops and the school) and the student pickup and drop-off times at the school and the stops. The intuition is that we would like to have short trips because of the savings in travel time and the fact that shorter trips could contribute to compatibility. The pickup and drop-off times, $PT_s, DT_s$, used in Equation (3) are calculated based on regression analysis. We use the results from Braca et al. (1997).

To account for the trip-to-school compatibility, we use an indicator function (1{$Condition$}) in Equation (4), which equals one if the condition is true and zero otherwise. Since we are minimizing the insertion cost, in Equation (4), the trip compatibility criterion is the total number of schools minus the schools that trip $j$ is compatible with after inserting stop $i$. This is the first consideration of compatibility in a fast heuristic algorithm. Clearly, the last stop of a trip and the travel time of the trip (defined in Equation (3)) will both affect the compatibility of the trip to other schools. And both the last stop and the travel time depend on the insertion location of a stop in that trip. In this paper, in order to calculate these costs, we augment the trips by inserting the stop in a position of the trip where the travel time of the augmented trip would be minimized. Some constraints like conservation of flow and sub-tour eliminations are not listed because they are implicitly guaranteed when the trips are generated under the insertion-based procedure. We explicitly try to enforce that other constraints such as the maximum ride time and capacity constraints do not get violated by adding a huge violation penalty to the weight between a stop and a trip if the addition of that stop would cause the trip to violate those conditions.

3.1.2.1 Solving the minimum cost matching problem and maintaining feasibility

The formulation of the minimum cost perfect matching problem is shown in Equations (5) – (8) with the insertion cost ($IC_{i,j}$) calculated using Equation (1). For the cases when $r \neq n$, we can add dummy stops or trips when needed to make it a perfect matching problem.

$$\text{Min } \sum_{i \in R} \sum_{j \in N} IC_{i,j} \cdot y_{i,j} \quad (5)$$
Subject to
$$\sum_{i \in R} y_{i,j} = 1, \forall j \in N + DummyTrips \quad (6)$$
$$\sum_{j \in N} y_{i,j} = 1, \forall i \in R + DummyStops \quad (7)$$
$$y_{i,j} \geq 0, \forall i \in R, j \in N \quad (8)$$

Objective (5) is to minimize the total sum of the insertion cost. Constraints (6) and (7) are enforcing that exactly one insertion happens for each pair of stop and trip. If an insertion involves the dummy stops or trips, it will be discarded. We use the Hungarian (Munkres) algorithm to solve the minimum cost matching problem. For PMCM, the minimum cost matching problem is to find $m$ assignments of stops to trips such that the total insertion cost is minimized where $m \leq \min(r, n)$ (before adding the dummy stop or trip). Each matching (assignment) indicates the insertion of one stop to one specific trip. All the stops and trips in these $m$ matches are distinguished. If these matches are feasible, the stops will be inserted into their corresponding trips. The trip set will be updated by these inserted trips, and these routed stops will be removed from the un-routed stop set. This previous step is step 5.2 in the MCM algorithm. The feasibility check (step 5.3 in the MCM algorithm) involves the vehicle capacity and maximum ride time constraints. An insertion is feasible only if both constraints are satisfied. For SMCM, only one stop with the minimum insertion cost will be inserted into the current trip being constructed. At the point when no feasible insertions exist for both PMCM and SMCM, a new trip is initialized by connecting the school to the furthest stop in the un-routed stop set. In PMCM, if all min(r, n) matches are feasible, all of these stops will be inserted simultaneously at one iteration compared to one-stop insertion per iteration for SMCM. Theoretically, PMCM could be faster than SMCM because it could be implemented in distributed settings and run in parallel. On the other hand, we experimentally show that there is no general rule to compare the solution quality of the sequential and parallel routing algorithms. In a not time-dependent application, it is desired to perform both PMCM and SMCM algorithms and select the one with the best solution. These sensitivity analyses are performed in section 4.1 and 4.2.

*3.1.3.Example*

A simple graphical example of PMCM with one school and 11 stops is shown in Figure 3. Assuming the initial number of trips equal to 3, in Figure 3 (a), stops are clustered into three clusters. Then, the furthest stops are selected for each cluster in Figure 3 (b). The trip set is initialized by adding the school stops and these furthest stops in Figure 3 (c). In Figure 3 (d), a minimum cost matching problem is solved, and three feasible matches are obtained. These three stops are inserted into their corresponding trips in Figure 3 (e). At Figure 3 (f), since no feasible insertion exists, a new trip (Trip 4) is initialized, and the algorithm continues.

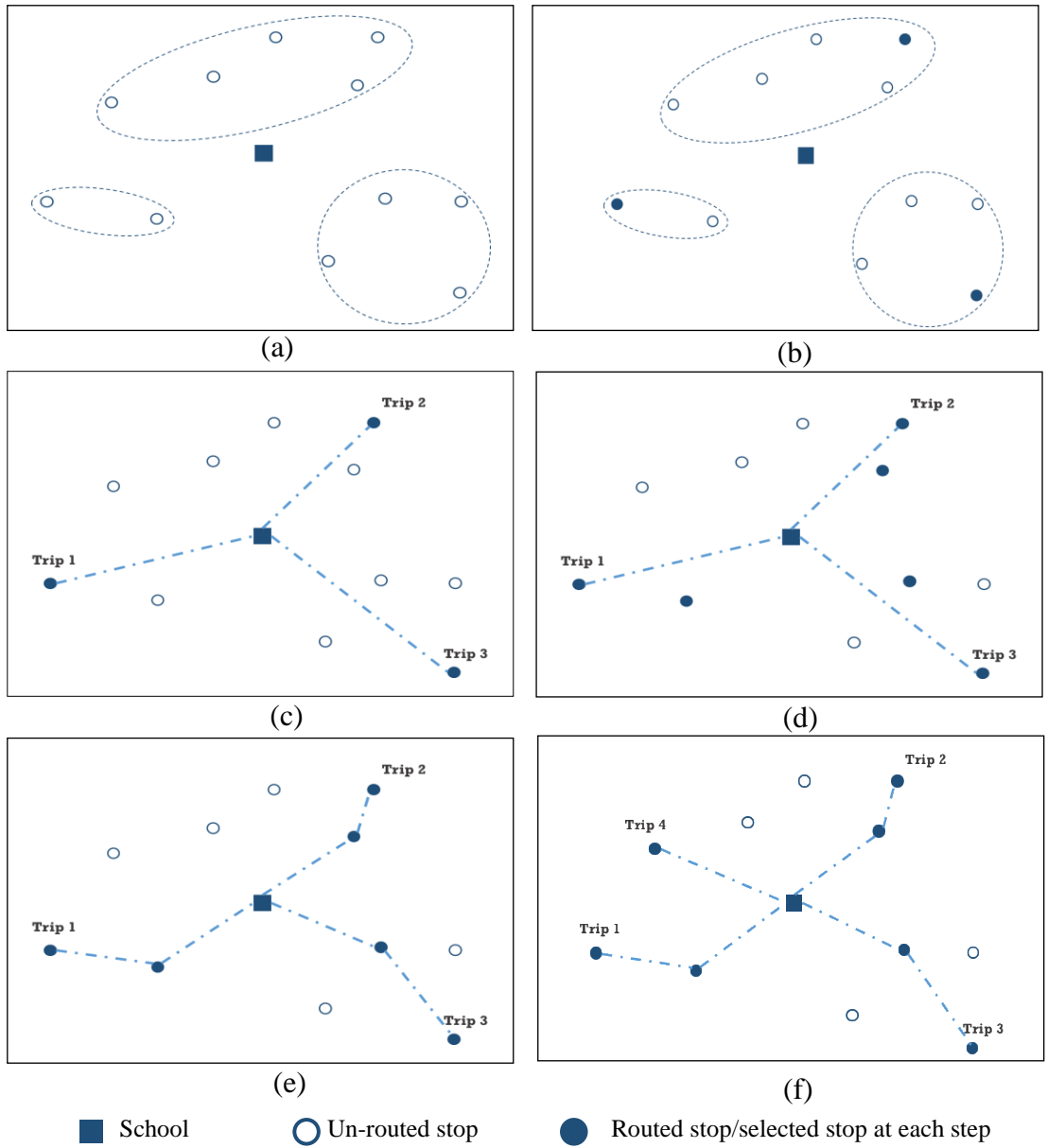

Figure 3 Graphical example of minimum cost matching routing algorithm

### 3.2. Post-Improvement Algorithm

The post-improvement (PI) algorithm is applied to the initial solution generated by the minimum cost matching (MCM) routing algorithm. It is a combination of Tabu Search, Simulated Annealing, and the chain exchange algorithm. The generic pseudo-code is shown in Algorithm 2 with notations from both Table 1 and Table 2.

Table 2 Notation for post improvement algorithm

| Notation | Description |
| --- | --- |
| $t$ | Temperature |
| $t_{end}$ | Terminal temperature |
| $t_{cool}$ | Temperature cooling rate |
| $it$ | Number of iteration that no changes occur |
| $it_{max}$ | Maximum number of iterations that no changes occur |
| $N$ | Trip set |
| $TL$ | Tabu list |
| $CR_p$ | Cost of remove for stop $p$ |
| $IS_p$ | Number of stops in the trip to which stop $p$ belongs |
| $IT_p$ | Travel time for the trip to which stop $p$ belongs |
| $IP_{s,p}$ | The closeness of stop $s$ with respect to stop p |
| $IQ_s$ | Number of students for the trip to which stop $s$ belongs |
| $IC_s$ | Trip compatibility of the trip to which stop $s$ belongs |
| $\epsilon$ | Random variables |
| $\beta_S$ | The coefficient of the number of stops in 'cost of remove' |
| $\beta_T$ | The coefficient of the travel time in 'cost of remove' and 'closeness' |
| $\beta_Q$ | The coefficient of the number of students in 'closeness' |
| $\beta_D$ | The coefficient of the distance in 'closeness' |
| $EL$ | Vertex exchange list |
| $NL$ | Neighborhood list |
| $Nnei$ | Maximum number of stops in the neighborhood list |
| $move(pl,p)$ | Move stop $p$ to the place right after stop $pl$ where p and pl belongs to different trips |
| $SC$ | Surrogate cost of a routing plan |
| $TN$ | Number of trips in a routing plan |
| $TC$ | Trip compatibility of a routing plan |
| $TT$ | The total travel time of a routing plan |
| $\gamma_N$ | The coefficient of the number of trips in surrogate cost |
| $\gamma_C$ | Coefficient of trip compatibility in surrogate cost |
| $\gamma_T$ | The coefficient of total travel time in surrogate cost |

There are two control variables in PI, the temperature *t* and number of iterations with no changes (*it*). These two control variables determine whether PI will perform local search (vertex exchange) or accept the current solution and terminate. The local search will be activated only if both of the following conditions are met: 1) temperature *t* is greater than or equal to the terminal temperature ($t_{end}$) and 2) that the number of iterations with no changes (*it*) is less than or equal to $it_{max}$.

**Algorithm 2 PI**

1. Solve the SBS using the initial routing plan, store as initial scheduling plan (ISP)
2. Initialize $t$, $it$, N, TL
3. **while** ($t \geq t_{end}$ and $it \leq it_{max}$) **do**
      3.1 construct vertex exchange list (EL)
      3.2 **for** (p in EL) **do**
         3.2.1 construct neighborhood list (NL[p]) of stop p
         3.2.2 **for** (pl in NL[p]) **do**
            A. **If** (*move(pl,p)* is feasible) **then**
              B. **If** (*move(p,pl)* is not in the Tabu list) **then**
                 C. **If** (*move(pl,p)* is acceptable at *t*) **then**
                    D. Implement *move(pl,p)*, update N, TL, $it = 0$, **Break**;
      3.3 $it$ += 1; $t = t$* t_cool
4. Solve the SBS using the new routing plan, store as new scheduling plan (NSP)
5. **If** (Number of buses of NSP < number of buses of ISP) **then**
      5.1 **Return** NSP
   **Else**
      5.2 **Return** ISP

*3.2.1. Vertex exchange list*

    The underlying mechanism for the vertex exchange operator is that when moving some stops from short trips to long trips, some short trips will become empty and then are deleted. As a result, the trip set will be reduced, and the route scheduling problem is more likely to come up with a solution with fewer buses. More accurately, other attributes like travel time and trip compatibility also contribute to the 'goodness' of a routing plan. The complete surrogate cost to evaluate the 'goodness' of a move will be discussed in section 3.2.3. The vertex exchange operator starts with constructing the vertex exchange list based on their 'cost of removal' as defined in Equation (9):

$$CR_p = \beta_S IS_p + \beta_T IT_p + \epsilon \tag{9}$$

where $IS_p, IT_p$ are the number of stops and travel time of the trip to which stop *p* belongs, and $\beta_S, \beta_T$ are the associated coefficients. The travel time is calculated the same way as in Equation (3) and it includes the pickup and drop-off time. The 'cost of removal' of a stop describes some measure of the negative likelihood of reaching a better routing plan after removing this stop. The vertex exchange list is the set of all stops arranged in the increasing order of their 'cost of removal'. The PI algorithm will try to move a stop in vertex exchange list sequentially to its neighbors such that some short trips will become empty and deleted. It is desired to first move the stops belonging to shorter trips (with respect to the number of stops and travel time). The short trips after removing these stops are more likely to be deleted. In this way, the total number of trips will likely decrease and the number of buses in the scheduling stage is more likely to be reduced. Because all the stops in the same trip will have the same number of stops and travel time, $\epsilon$ is used to introduce some randomness to help to determine the sequence of these stops in the exchange list.

*3.2.2. Neighborhood list*

    For each stop *p* in the vertex exchange list, its neighborhood list (NL) contains the stops

that are close to stop *p* such that stop *p* has a higher probability to be inserted after them. The stops in the neighborhood list are sorted in the ascending order of the 'closeness' (IP) as follows:

$$IP_{s,p} = \beta_Q IQ_s + \beta_T IT_s + \beta_D D_{s,p} \quad \forall s \in Stops \backslash Stops_p \tag{10}$$

where $IQ_s, IT_s, D_{s,p}$ are the number of students, the travel time of the trip to which stop *s* belongs, and the distance between stop *s* and *p*, and $\beta_Q, \beta_D, \beta_T$ are the associated coefficients. Equation (10) states that stop *p* a has a higher probability to be inserted after a stop *s* which is close to stop *p* and belongs to the trip that has few students and has a short travel time. It is inefficient to check the feasibility of every pair of vertex move and it is undesirable to insert stop *s* into a faraway trip even if it is feasible. Therefore, the size of the neighborhood is limited by Nnei (<Nst) and all the stops in the neighborhood will be examined sequentially until one move is feasible and acceptable or until we reach the end of the list. The neighborhood does not contain the stops that are in the same trip with stop *p* since the main goal of the vertex exchange is to move stop *p* to another trip such that the old trip of stop *p* might be deleted.

For each stop *pl* in the neighborhood list (NL) of stop *p*, PI checks whether move (p, pl) is in the Tabu list and that whether move (pl, p) is feasible. The feasibility check consists of the maximum ride time and vehicle capacity constraints. PI only considers the feasible move of which its reversed move is not accepted before. This idea is from Tabu Search and implemented to avoid frequent reversed moves. In the PI algorithm, the acceptance of a move is based on the Simulated Annealing criteria where a move is still acceptable even if it yields a worse cost function. It means that move (pl, p) and move (p, pl) may be both acceptable. One consequence is that once a move is removed from TL, it is very likely to be implemented again even if it yields a worse cost function. Hence, the maximum forbidden iteration should be large enough to prevent the frequent removal of forbidden moves.

*3.2.3. Surrogate cost*

As explained in the introduction, the real estimate of the goodness of a routing plan is the number of buses used after solving a route scheduling problem, which is unknown during trip generating stage. The surrogate cost (SC) is used to assess the goodness of the routing plan and the goodness of a move without solving route scheduling problem. We adopt the proposed surrogate cost function from Shafahi et al. (2017b). This is presented in Equation (11):

$$SC = \gamma_N TN - \gamma_C TC + \gamma_T TT \tag{11}$$

where $TN, TC, TT$ are the total number of trips, trip compatibilities and travel time of a routing plan with coefficients $\gamma_N, \gamma_C, \gamma_T$. Total trip compatibility is the sum of trip-to-trip compatibility which is similar to Equation (4) except that the trip-to-school compatibility is replaced by trip-to-trip compatibility. The total travel time is the sum of the travel time and the pickup and drop-off time as calculated in Equation (3). The surrogate cost implies that a routing plan is a good one if the number of trips and total travel time is small while the trip compatibility is as large as possible. This is the instance that trip-compatibility appears in our proposed heuristic algorithm. Still, Equation (11) is just a surrogate estimate of the goodness of a routing plan, it is possible that a routing plan with smaller SC needs more buses than another one with higher SC. PI adopts the acceptance criterion from Simulated Annealing in Equation (12) to help to jump local optima by introducing some randomness.

$$P(accept\ a\ move) = 1/(1 + e^{(SC_{new} - SC_{old})/t}) \qquad (12)$$

Equation (12) ensures that when the new surrogate cost after a move is smaller than the old one (which is preferred), this move will be accepted. On the other hand, if the new surrogate cost is larger than the old one (which is undesired), the probability of accepting this move is small if the difference between two surrogate costs is large and the temperature is small. PI will be iteratively implemented until either the maximum number of iterations without change reaches or the temperature drops below the terminal temperature.

*3.2.4. Example*

An illustrative example of PI is presented in Figure 4. Assume that at the current iteration, the parameters are $t = 10, it = 2, TL = \{(1,4)\}$. There are three trips connecting five stops for school $a$. In Figure 4 (a), PI first constructs the vertex exchange list (*EL*) as $\{5,3,1,4,2\}$. For the first stop (5) in the *NL*, find its neighborhood list $NL[5]=\{4, 2, 1\}$ with $Nnei = 3$. Starting from the first move(4,5), it satisfies all conditions (A: feasibility, B: its reverse mode is not in the Tabu list, C: it is acceptable given the surrogate cost estimate). Then stop 5 is removed from trip 3 to the position right after stop 4 in trip 2. Trip 3 becomes empty and is deleted while trip 2 is augmented to $\{3,4,5\}$. This iteration is terminated, and control parameters are updated: $t = 10 \times 0.9 = 9; it = 0; TL = \{(1,4), (4,5)\}$. Note that in this iteration, one stop is moved, and one trip is deleted, the *it* (number of iterations that no change occurs) count is reset to zero.

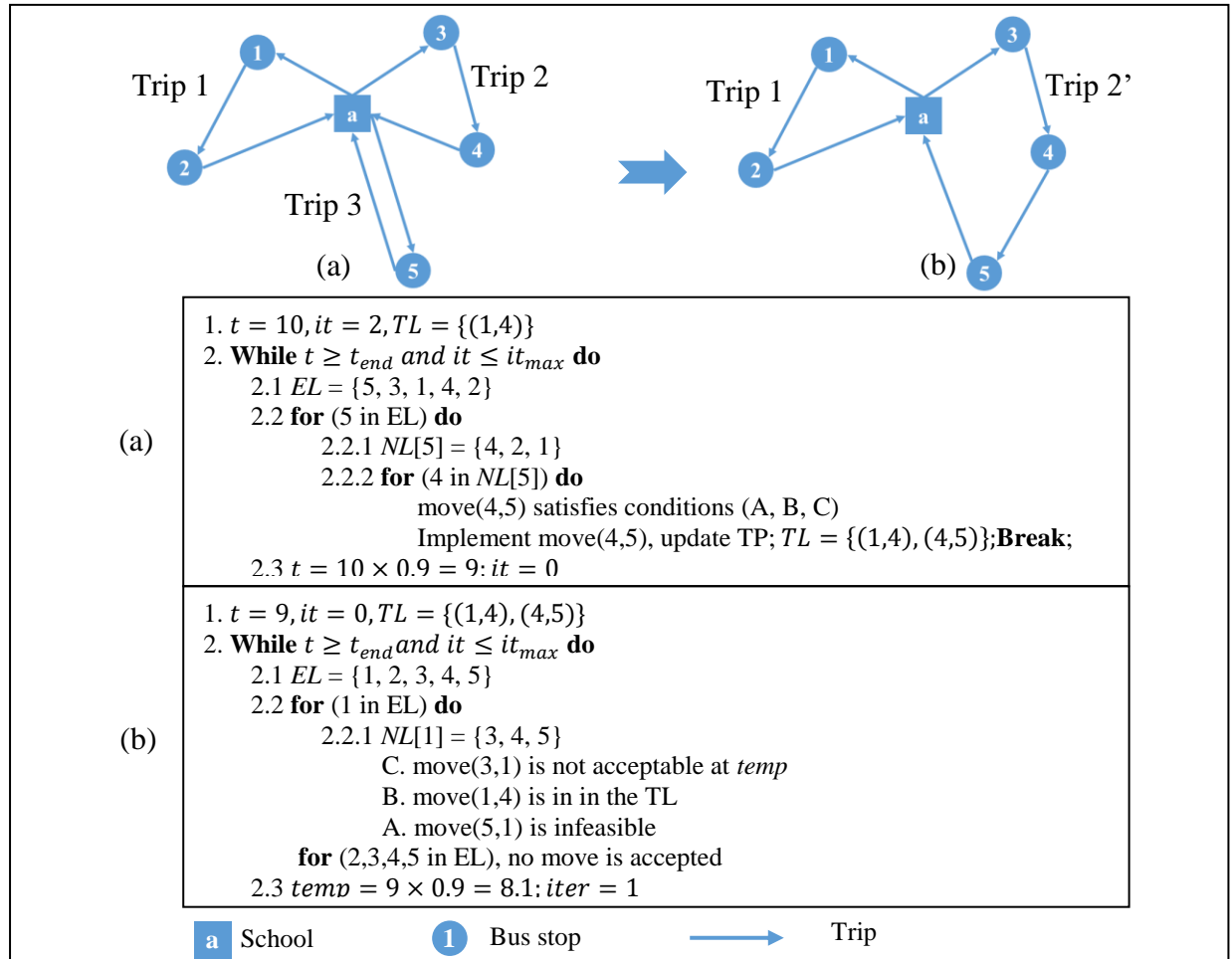

Figure 4 Illustration of Post Improvement Algorithm

At the next iteration, PI re-constructs the vertex exchange list (*EL*) as {1, 2, 3, 4, 5}. For the first stop, 1, PI finds its neighborhood list *NP*[1]={3,4,5}. PI finds none of the moves between stop 1 and stops 3, 4, and 5 is feasible or acceptable. PI then tries to remove stop 2. The process is repeated, and no move is accepted for stop 2, 3, 4, and 5. This iteration terminates without any change. The temperature cools down, and iteration count is incremented by one ($t = 9 \times 0.9 = 8.1; it = 1$). PI continues for the next iteration. Notice that due to the randomness of constructing vertex exchange list and the randomness of accepting criterion, some unacceptable moves might become acceptable at different iterations. Therefore, it is worthy to continue the search even if no change occurs at one iteration.

## 4. Computational Results

We conduct different experiments. First, we conduct experiments on randomly generated examples to compare the solution quality and efficiency of the proposed MCMP heuristic with the exact method and to compare the performance of the Sequential Minimum Cost Matching routing (SMCM) with the Parallel Minimum Cost Matching routing (PMCM). Then, a preliminary experiment and a sensitivity analysis experiment are conducted on a benchmark problem. In the end, we use PMCM to solve all the benchmark problems from Park et al. (2012). We compare the state-of-the-art solutions on the benchmark problems with the solutions from our PMCM (number of buses) algorithm both with and without the post-improvement step under the best meta-parameter settings obtained from the sensitivity analysis study.

As mentioned before, routing is only a part of the school bus planning problem; the main criterion is the number of buses obtained from solving the scheduling problem. In all the following experiments, after generating routing plans from different algorithms, all the routing plans were treated as the input for the scheduling problem. The scheduling problems were solved using Kim et al. (2012)'s formulation, which was an assignment problem under the assumption of the homogeneous fleet and fixed trip start times. The number of buses (NB) is the main criterion used to evaluate the goodness of a routing plan. Thanks to the optimality of the scheduling problems, the number of buses outputted by solving the scheduling problem is an accurate and fair criterion for the different routing plans. The MCM and PI algorithms were implemented in Python 2.7., after setting up the problem as a minimum cost matching problem, the minimum cost matching problem was solved using SciPy optimization tool, and the scheduling problem was solved using Gurobi. All the experiments were run on a computer with i7 CPU 870 @ 2.93 GHz and 8GB RAM.

4.1. Randomly generated problems
*4.1.1.Random problem data generation*
A set of test problems were randomly generated with various sizes of schools and stops. For each test problem, all nodes (including stops and schools) were assumed to be located within a 2-dimensional square with the length of 40 miles (211,200 ft). The locations of all nodes are designated using their longitude and latitude, which were both randomly generated as a uniformly distributed random variable between 0 and 211220. Then, we applied the K-means algorithm with the number of clusters equal to the number of schools and found the centroid of each cluster. The closest node to each cluster centroid was selected to be the school location, and the rest of the nodes were treated as bus stops and were assigned to their closest school. The number of students at each stop was randomly generated as a uniformly distributed random variable between 1 and 20. The bus capacity was assumed to be 66 and the maximum ride time was assumed to be 90 minutes. The bell times for the schools are random uniform variable between 12:00 PM and 16:00 PM and they are integer multiple of 15 minutes (i.e., 2:00 PM, 3:15 PM). The distance is Euclidean. The bus runs at a constant speed of 20 miles per hour. The parameters were set to be $\alpha_Q = \alpha_C = \alpha_T = 1$ for both SMCM and PMCM. Pickup (Equation (24)) and drop off (Equation (25)) time came from the regression model developed by Braca et al. (1997).

$$\text{PT}_k = 29.0 + 1.9n_s, \forall k \in \text{Schools} \quad (24)$$
$$\text{DT}_i = 19.0 + 2.6q_i, \forall i \in \text{Stops} \quad (25)$$

where $\text{PT}_k, \text{DT}_i$ are the pickup and drop off durations at schools and stops in seconds, and $n_s, q_s$ are the number of student at the schools and the stops. Note that the original regression from Braca et al. (1997) was for the morning trips where students were picked up from stops and dropped off at schools. In our afternoon trip setting, we consider the pickup time from stops to be similar to the drop-off time at stops and the drop-off time at the school equal to the pickup time from schools.

*4.1.2. Running time of SMCM and PMCM and exact method*

The first experiment sheds light on the necessity of fast heuristic to solve single-school bus routing problems. This experiment tests how the running time of the exact method and MCM would increase in response to the increase of the problem size. In this experiment, the number of schools is set to two and the number of stops increases from 20 to 60. The approaches used are:

1) EXA: Directly solve the single school routing problem (from Shafahi et al., 2017b) to optimality by commercial solver;
2) SMCM: Solve the single school routing problem by sequential minimum cost matching algorithm without post improvement;
3) PMCM: Solve the single school routing problem by parallel minimum cost matching algorithm without post improvement.

To obtain NB, the scheduling problem is solved given the routing plans generated by these methods. The result is shown in Table 3 and Figure 5. In small cases (scenario 1, 3 and 4), both SMCM and PMCM can find solutions as good as the exact method (with respect to the number of buses) in much shorter time (as low as 0.16% in scenario 4). The difference arises when the problem size gets bigger. The EXA can find solutions using a fewer number of buses than the PMCM and SMCM at the expense of huge running time (scenario 5 and 7). The difference between the SMCM and PMCM is also observed. Out of eight scenarios, the SMCM finds the solutions using the same number of buses as the EXA in five scenarios where the PMCM only has four. Still, the solutions from the PMCM and SMCM are both quite good. The SMCM can find solutions using at most one more buses than the exact method, and the PMCM only uses at most two more buses. The total travel time from the PMCM and SMCM is higher than the EXA. But the travel time is much less important than the number of buses. Such difference can be ignored in large scenarios where the exact method is nearly unsolvable. The running time of the exact method becomes a critical problem with the increase of the problem size. The running time for heuristic barely increases with the increase of the problem size. But the running time for the exact methods increases exponentially. With such exponential trend, the problem can easily become unsolvable using exact methods in fact, if the number of stops is increased to 100, even for this still small size problem, the exact method cannot find the optimal solution within 5 days. This showcases the necessity of fast heuristics for solving large single-school problems.

Table 3 Solution summary for exact method, PMCM and SMCM

| Scenario | | 1 | 2 | 3 | 4 | 5 | 6 | 7 | 8 |
|---|---|---|---|---|---|---|---|---|---|
| # of schools | | 2 | 2 | 2 | 2 | 2 | 2 | 2 | 2 |
| # of stops | | 20 | 24 | 28 | 32 | 36 | 40 | 50 | 60 |
| EXA | NB | **4*** | **2*** | **3*** | **4*** | **3*** | **4*** | **5*** | **9*** |
| | NT | 4 | 5 | 5 | 6 | 6 | 8 | 8 | 10 |
| | TT | 313 | 327 | 375 | 393 | 433 | 540 | 488 | 721 |
| | **RT** | **2.36** | **4.60** | **155** | **191** | **143** | **610** | **1654** | **4464** |
| PMCM | NB | **4*** | **3** | **3*** | **4*** | **5** | **5** | **7** | **9*** |
| | NT | 4 | 5 | 5 | 6 | 6 | 9 | 8 | 10 |
| | TT | 446 | 468 | 474 | 535 | 618 | 724 | 762 | 921 |
| | **RT** | **0.07** | **0.10** | **0.68** | **0.61** | **0.11** | **0.50** | **1.02** | **0.22** |
| SMCM | NB | **4*** | **3** | **3*** | **4*** | **4** | **4*** | **6** | **9*** |
| | NT | 4 | 5 | 5 | 6 | 7 | 8 | 8 | 10 |
| | TT | 443 | 488 | 566 | 560 | 660 | 763 | 806 | 1013 |
| | **RT** | **0.13** | **1.45** | **1.84** | **0.30** | **0.26** | **0.48** | **0.63** | **0.79** |

**Note:** * minimum number of buses; NB: Number of buses; NT: Number of trips; TT: Total travel time (minutes); RT: Running time (second)

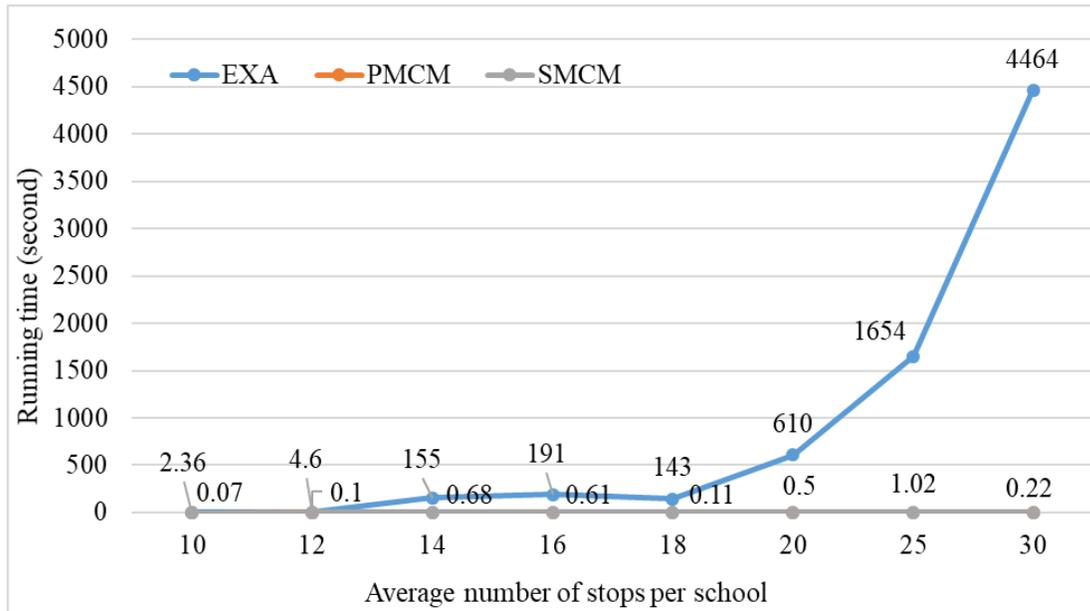

Figure 5 Running time comparison between the EXAA, PMCM, and SMCM

*4.1.3. SMCM and PMCM on large size problems*

After showing the heuristic algorithms are much faster than the exact method on small size problems, in this subsection, want to answer the question of whether the heuristic algorithm still efficient to solve large size problems. The test problems are generated using a similar method described in section 4.1.1 except that the problem size goes up to 300 schools and 6000 stops. The result of the comparison between SMCM and PMCM is shown in Table 4. Out of 11 test scenarios, SMCM found a better result (in terms of fewer number of buses) six times compared to two times when PMCM found a better result. However, the difference was not significant; the differences were no bigger than 10%. Also, there were three scenarios that SMCM and PMCM found the results using the same number of buses. Therefore, with respect to the minimum number of buses, SMCM had a slightly better result than PMCM, but it is not significant. When considering the efficiency, PMCM outperforms SMCM. The running time for PMCM is around half of the running time for SMCM especially for larger cases where the initial fixed running time for loading the problem had a smaller impact on the overall running time. Both SMCM and PMCM were pretty efficient. The largest problem with 300 schools, 6000 stops, and over 1000 trips can be solved within 6 minutes. In the next subsection, we further compare these two algorithms on more realistic benchmark problems.

Table 4 Comparison between SMCM and PMCM

| Scenario | # of schools | # of stops | SMCM | | | PMCM | | | Difference | |
|---|---|---|---|---|---|---|---|---|---|---|
| | | | NT[1] | NB[2] | RT[3] | NT | NB | RT | Diff[2] | %[3] |
| 1 | 2 | 60 | 15 | **15** | 0.53 | 14 | **14**[*] | 0.52 | 1 | 7% |
| 2 | 4 | 120 | 24 | **18**[*] | 0.74 | 28 | **19** | 0.91 | 1 | 6% |
| 3 | 6 | 180 | 34 | **21**[*] | 1.15 | 37 | **23** | 0.63 | 2 | 10% |
| 4 | 8 | 240 | 39 | **26** | 1.55 | 42 | **26** | 0.75 | 0 | 0% |
| 5 | 10 | 300 | 49 | **32** | 2.14 | 53 | **32** | 0.96 | 0 | 0% |
| 6 | 20 | 600 | 102 | **52**[*] | 4.89 | 111 | **56** | 2.34 | 4 | 8% |
| 7 | 50 | 1000 | 181 | **71**[*] | 10.83 | 191 | **73** | 6.00 | 2 | 3% |
| 8 | 100 | 2000 | 364 | **116** | 36.34 | 384 | **116** | 21.61 | 0 | 0% |
| 9 | 150 | 3000 | 528 | **132**[*] | 89.59 | 570 | **139** | 48.47 | 7 | 5% |
| 10 | 200 | 4000 | 734 | **176** | 167.96 | 788 | **174**[*] | 88.54 | 2 | 1% |
| 11 | 300 | 6000 | 1065 | **201**[*] | 376.27 | 1151 | **209** | 197.30 | 8 | 4% |

[*]: best result with respect to the minimum number of buses; [1]NT: Number of trips; [2]NB: Number of buses; [3]RT: running time for routing problem (seconds); [2] $\text{Diff} = |\# \text{ buses (SMCM)} - \# \text{ buses (PMCM)}|$; [3]% = Diff / $\min\{\# buses\ (SMCM), \# buses\ (PMCM)\} \times 100\%$.

4.2. Preliminary experiment for benchmark problems

The benchmark problems solved in this paper were developed by Park et al. (2012). There are two sets of test scenarios: 1) RSRB, where stops and schools are randomly distributed and 2) CSCB, where stops and schools are gathered together. There are eight test problems in RSRB up to 100 schools, 2000 stops and 31939 students and 16 test problems in CSCB up to 100 schools, 2000 stops and 27945 students. The maximum ride time (MRT) was set to be either 2700 seconds (45 minutes) or 5400 seconds (90 minutes). All 24 test problems were solved using two different maximum ride times, making it 48 separate test cases in total. The bus fleet is assumed to be homogeneous with the fixed capacity of 66. The distance is assumed to be grid (Manhattan) distance with a constant bus speed of 20 miles per hour. The pickup and drop off time were the same as those in Equations (24) and (25). All Park et al. (2012)'s results were adopted using the corrected version of Aug 10th, 2012.

The preliminary experiment was made between SMCM and PMCM without Post Improvement algorithm using the same naïve parameters setting ($\alpha_Q = \alpha_C = \alpha_T = 1$). The experiments were limited to the first five scenarios for both RSRB and CSCB under the maximum ride time (MRT) of 4500 seconds. The result is summarized in Table 5. The major criterion is the same as before, the number of buses (NB) and the efficiency (running time, RT). The comparison was made between the solution from Park et al. (2012) (abbreviated as Park) with single-load and the solution that Park et al. (2012) reported using Braca et al. (1997) algorithm (abbreviated as Braca). It can be seen that SMCM and PMCM had very similar results for the number of buses and the difference was within 4 buses. When compared to the solution from Braca and Park, both SMCM and PMCM can find better results than the existing methods for about half of the cases even under the naïve parameter setting. This shows the potential of MCM algorithm. The running time for MCM was higher than Park because 1) MCM was coded in Python while Park algorithm was coded in C, which is inherently faster than Python; and 2) our code is not optimized since the running time is not a critical issue for a planning problem. Still, it can be seen that under the same coding structure and language, PMCM was much faster than SMCM. Therefore, the detailed computational tests in the next section use PMCM due to its efficiency and the somewhat similar solution quality as SMCM.

Table 5 Preliminary experiment of SMCM and PMCM on benchmark problems

| Scenario | # of schools | # of stops | Braca NB[1] | Park | | SMCM | | | PMCM | | |
|---|---|---|---|---|---|---|---|---|---|---|---|
| | | | | NB | RT[2] | NT[3] | NB | RT | NT | NB | RT |
| RSRB01 | 6 | 250 | **31** | **31** | 0.5 | 60 | 32 | 2.9 | 56 | **31** | 0.6 |
| RSRB02 | 12 | 250 | **29** | 30 | 0.5 | 68 | **27**[+*] | 1.4 | 71 | 31 | 0.9 |
| RSRB03 | 12 | 500 | **61** | **61** | 0.8 | 126 | **52**[+*] | 6.2 | 124 | **49**[+*] | 2.2 |
| RSRB04 | 25 | 500 | 60 | **56** | 0.9 | 140 | **56**[+] | 3.7 | 138 | **54**[+*] | 2.4 |
| RSRB05 | 25 | 1000 | **86** | 106 | 1.5 | 234 | 92[*] | 24.7 | 235 | 92[*] | 4.8 |
| CSCB01 | 6 | 250 | **33** | 35 | 0.5 | 69 | 34[*] | 4.3 | 69 | 34[*] | 0.9 |
| CSCB02 | 12 | 250 | **26** | 27 | 0.5 | 67 | 30 | 3.2 | 66 | 28 | 0.9 |
| CSCB03 | 12 | 500 | **50** | 52 | 0.7 | 133 | 58 | 8.7 | 127 | 53 | 2.9 |
| CSCB04 | 25 | 500 | 67 | 57 | 0.9 | 150 | **54**[*+] | 4.8 | 159 | **54**[*+] | 3.1 |
| CSCB05 | 25 | 1000 | 134 | 115 | 1.6 | 291 | **119**[+] | 31.7 | 308 | **116**[+] | 6.6 |

**Note**: [1]NB: Number of buses; [2]RT: running time without route scheduling problem (seconds); [3]NT: Number of trips; [*]: if the result was better than Park et al. (2012); [+]: the result was better than Braca et al. (1997).

### 4.3. Benchmark problems
#### 4.3.1. Sensitivity analysis

Clearly, parameters have a significant impact on the algorithms. In this section, we experiment with different values of the meta-parameters of both the minimum cost matching (MCM) and the post-improvement (PI) algorithm. A way of choosing a good combination of these meta-parameters is conducting sensitivity analysis. Due to the huge combination of different parameters in MCM and PI, only a set of parameters candidates were tested. The sensitivity analysis shown in Table 6 was conducted based on RSRB01 with MRT=2700 seconds. The most influential meta-parameters in MCM (ParaMCM) were coefficient for the remaining vehicle capacity $(\alpha_Q)$, trip compatibility $(\alpha_C)$ and travel time $(\alpha_T)$ for evaluating insertion cost in Equation (1). Eight sets of meta-parameters for ParaMCM were tested in the form of $(\alpha_Q:\alpha_C:\alpha_T)$. And the most important meta-parameters in post improvement (ParaPI) were the coefficients for the total number of trips $(\gamma_N)$, trip compatibilities $(\gamma_C)$ and travel time $(\gamma_T)$ in the surrogate cost estimation (Equation (11)). Four sets of meta-parameter values were examined in the shape of $(\gamma_N:\gamma_C:\gamma_T)$. In total, 32 different combinations of meta-parameter sets were examined.

The criterion was still the number of buses (NB). Under each meta-parameter set test, the solution from pure PMCM and solution from PMCMPI (implementing PI after PMCM) were both reported. Other parameters were set to be $\beta_s = 10000, \beta_Q = 100, \beta_D = 10, \beta_T = 1, t_{initial} = 100 \times n, t_{end} = 10, t_{cool} = 0.9, it_{\max} = 10, \epsilon \sim Uni(0,100), Nnei = 20$. Note that the trip compatibility calculation (Equation (4)) requires the fixed scalar school bell time such that the start time of the trip equals to the dismissal time of the school. In the benchmark problem from Park et al. (2012), the school dismissal (or for morning bell) time is a time window and the start of the trip is a variable within that time window. In our implementation of MCMPI, we assumed that the start of each school equals to the early school bell time and the trip compatibility is a rough estimate of the real trip compatibility. The actual trip start time is calculated in the process of solving the route scheduling problem.

Table 6 Sensitivity analysis of MCM and PI

| NB[2] | | ParaPI[1] | | | | | | | |
|---|---|---|---|---|---|---|---|---|---|
| | | (10000:1250:1) | | (1000:125:1) | | (100:125:1) | | (10:10:1) | |
| | | PMCM[3] | PI[4] | PMCM | PI | PMCM | PI | PMCM | PI |
| Para MCM[5] | (300:20:1) | 50 | 34 | 49 | 34 | 51 | 36 | 50 | 35 |
| | (200:100:1) | 38 | 33 | 39 | 33 | 38 | 33 | 39 | 34 |
| | (100:100:1) | 36 | 35 | 34 | 34 | 35 | 35 | 36 | 35 |
| | (200:10:1) | 33 | **31*** | 37 | 33 | 37 | 33 | 38 | 35 |
| | (100:10:1) | 33 | 34 | 34 | 34 | 33 | 34 | 35 | 35 |
| | (1:1:1) | 34 | 34 | 35 | 35 | 35 | 35 | 33 | 33 |
| | (10:10:1) | 34 | 34 | 34 | 34 | 34 | 34 | 34 | 34 |
| | (10:100:1) | 35 | 35 | 34 | 34 | 34 | 34 | 34 | 34 |

**Note**: [1]parameters for PI in the form of $(\gamma_N:\gamma_C:\gamma_T)$, [2]NB: number of buses; [3]PMCM: number of buses of the solution generated by PMCM; [4]PI: number of buses of the solution generated by PMCM followed by PI; [5]ParaMCM: parameter of MCM in the form of $(\alpha_Q:\alpha_C:\alpha_T)$; [*]: best solution

By changing the meta-parameters, MCM and PI perform quite differently. For example, under the same ParaPI ($\gamma_N:\gamma_C:\gamma_T = 100:125:1$), changing ParaMCM will lead to the solution changing from 33 buses to 51 buses. On the other hand, by changing ParaPI, the number of buses can be saved with respect to PMCM will also be changed. Because the parameters for MCM and PI are not independent, the cross-examination is a logical way of choosing an appropriate combination of parameters. Table 6 shows that the best solution (with 31 buses) was obtained when $\alpha_Q:\alpha_C:\alpha_T = 200:10:1$ and $\gamma_N:\gamma_C:\gamma_T = 10000:1250:1$. This set of parameters was used in all other test scenarios.

*4.3.2.Experiment setup*

PMCM was adopted to generate trips for all test scenarios compared to the solution reported by Park et al. (2012). Then, PI was implemented to hopefully further improve the solution from PMCM. There are three major sets of comparisons: 1) solutions from PMCM are compared to the existing solutions (Park et al., 2012), for both difference in the number of buses and relative bus saving percentage; 2) solutions from PMCMPI are compared to the existing solutions; and 3) performance of PI, which is the improvement of PMCMPI with respect to the solutions from PMCM. The results of these three sets of comparisons are summarized Table 7 and Table 8. Both PMCM and PMCMPI were run for ten iterations, and only the best result was reported. Because Park et al. (2012) reported better results than Braca et al. (1997) for almost all cases, in this paper, we only compare the number of buses from PMCM and PMCMPI to those reported by Park et al. (2012).

*4.3.3.Result*

    The improvement of PMCM and PI with respect to Park et al. (2012) (abbreviated as Park) can be observed in Table 7 (where maximum ride time is set to be 2700 seconds) and Table 8 (maximum ride time is set to be 2700 seconds). Among 24 test scenarios, PMCM found better or equal results than Park 20 times. The maximum bus saving ratio of PMCM with respect to Park was 13%. This saving was more significant for large cases like CSCB07 where 201 buses can be reduced to 175 buses. Thirteen percentage bus saving, in this case, equals 26 buses. The solution from PMCMPI was even better than PMCM. Once the post-improvement was added (PMCMPI), Park algorithm only outperforms PMCMPI in one scenario (CSCB11). The highest saving ratio can reach 19% compared to Park at CSCB07, and the highest number of buses saved reaches 27, which is a significant improvement. On average, PMCM can save 5.7 buses which roughly equals 5% compared to Park. Moreover, by applying PI, such saving went to 7.3 buses, which roughly equals 7% of the total number of buses. PI can improve the solution from PMCM by additional 2%. Similar results can be seen with MRT=5400 cases and the bus saving (exact number and percentage) is shown in Figure 6.

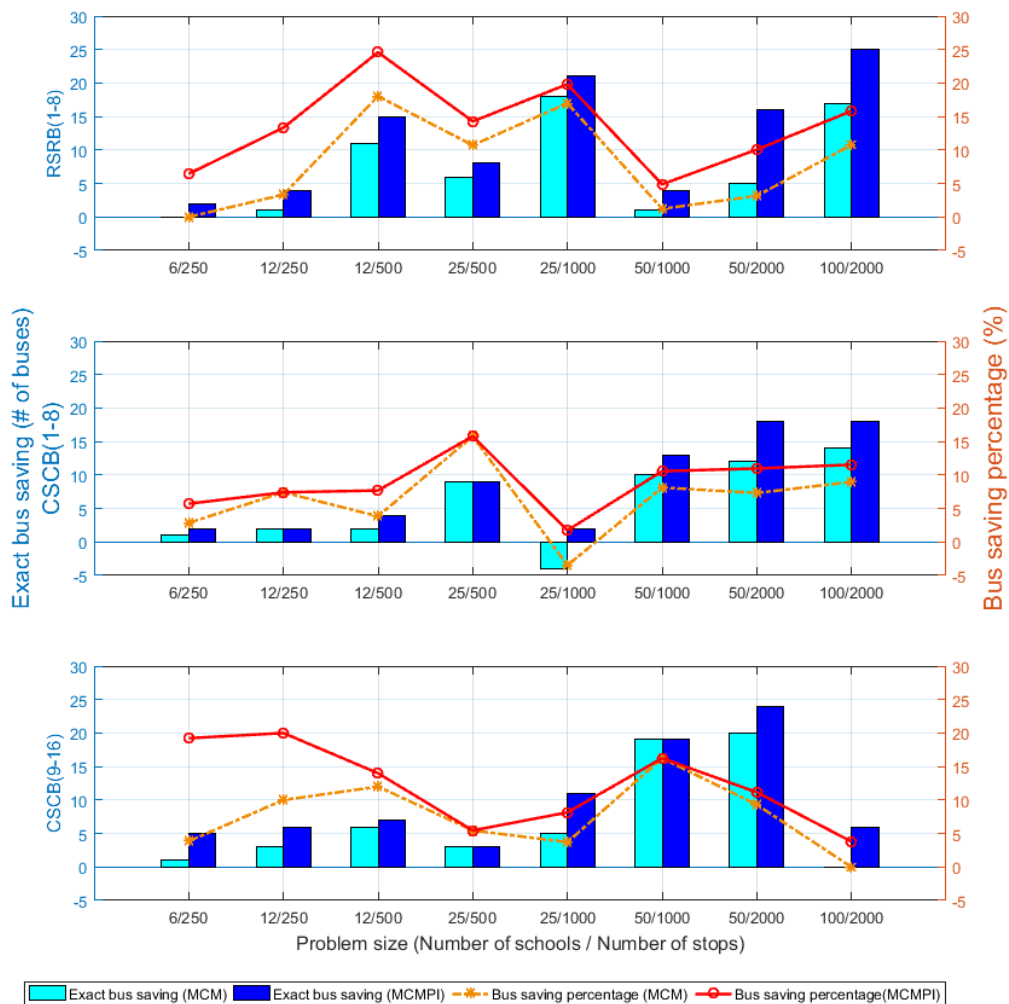

Figure 6 Bus saving compared to Park (MRT=5400 seconds (i.e., 90 minutes))

Table 7 Computation result of PMCM and PMCMPI for benchmark with MRT=2700

| Scenario | # of schools | # of stops | Park | | | PMCM | | | PMCMPI | | | PMCM2Park[4] | | PMCMPI2Park[7] | | PI2PMCM[8] | |
|---|---|---|---|---|---|---|---|---|---|---|---|---|---|---|---|---|---|
| | | | NT[1] | NB[2] | RT[3] | NT | NB | RT | NT | NB | RT | NB[5] | %[6] | NB | % | NB | % |
| **RSRB01** | 6 | 250 | 77 | **35** | 0.6 | 73 | **33** | 1.3 | 71 | **31** | 52.3 | 2 | 6% | 4 | 11% | 2 | 6% |
| **RSRB02** | 12 | 250 | 88 | **32** | 0.5 | 85 | **30** | 1.3 | 85 | **29** | 35.8 | 2 | 6% | 3 | 9% | 1 | 3% |
| **RSRB03** | 12 | 500 | 154 | **66** | 0.8 | 156 | **62** | 2.5 | 158 | **59** | 66.7 | 4 | 6% | 7 | 11% | 3 | 5% |
| **RSRB04** | 25 | 500 | 192 | **68** | 1 | 192 | **67** | 2.7 | 192 | **67** | 78.0 | 1 | 1% | 1 | 1% | 0 | 0% |
| **RSRB05** | 25 | 1000 | 315 | **124** | 1.6 | 303 | **113** | 10.6 | 304 | **112** | 143.7 | 11 | 9% | 12 | 10% | 1 | 1% |
| **RSRB06** | 50 | 1000 | 366 | **103** | 1.7 | 349 | **92** | 11.2 | 351 | **90** | 122.9 | 11 | 11% | 13 | 13% | 2 | 2% |
| **RSRB07** | 50 | 2000 | 637 | **185** | 2.6 | 599 | **181** | 38.1 | 593 | **179** | 168.0 | 4 | 2% | 6 | 3% | 2 | 1% |
| **RSRB08** | 100 | 2000 | 807 | **178** | 3.1 | 775 | **159** | 54.6 | 778 | **156** | 269.8 | 19 | 11% | 22 | 12% | 3 | 2% |
| **CSCB01** | 6 | 250 | 85 | **39** | 0.5 | 83 | **37** | 2.1 | 82 | **36** | 16.1 | 2 | 5% | 3 | 8% | 1 | 3% |
| **CSCB02** | 12 | 250 | 96 | **33** | 0.5 | 96 | **33** | 1.5 | 95 | **33** | 19.0 | 0 | 0% | 0 | 0% | 0 | 0% |
| **CSCB03** | 12 | 500 | 176 | **66** | 0.8 | 181 | **64** | 5.6 | 179 | **62** | 38.7 | 2 | 3% | 4 | 6% | 2 | 3% |
| **CSCB04** | 25 | 500 | 214 | **72** | 1 | 219 | **66** | 3.9 | 215 | **66** | 35.5 | 6 | 8% | 6 | 8% | 0 | 0% |
| **CSCB05** | 25 | 1000 | 398 | **135** | 1.8 | 390 | **137** | 14.3 | 393 | **134** | 79.6 | -2 | -1% | 1 | 1% | 3 | 2% |
| **CSCB06** | 50 | 1000 | 471 | **138** | 2 | 458 | **140** | 11.7 | 454 | **137** | 120.0 | -2 | -1% | 1 | 1% | 3 | 2% |
| **CSCB07** | 50 | 2000 | 724 | **201** | 2.9 | 730 | **175** | 87.8 | 725 | **174** | 214.3 | 26 | 13% | 27 | 13% | 1 | 1% |
| **CSCB08** | 100 | 2000 | 840 | **183** | 3.4 | 841 | **171** | 87.6 | 839 | **170** | 244.6 | 12 | 7% | 13 | 7% | 1 | 1% |
| **CSCB09** | 6 | 250 | 82 | **32** | 0.4 | 80 | **27** | 1.3 | 81 | **26** | 17.1 | 5 | 16% | 6 | 19% | 1 | 4% |
| **CSCB10** | 12 | 250 | 116 | **33** | 0.6 | 112 | **33** | 1.7 | 111 | **32** | 29.9 | 0 | 0% | 1 | 3% | 1 | 3% |
| **CSCB11** | 12 | 500 | 158 | **58** | 0.7 | 164 | **62** | 7.7 | 163 | **60** | 41.6 | -4 | -7% | -2 | -3% | 2 | 3% |
| **CSCB12** | 25 | 500 | 181 | **77** | 1 | 198 | **76** | 6.9 | 198 | **75** | 37.1 | 1 | 1% | 2 | 3% | 1 | 1% |
| **CSCB13** | 25 | 1000 | 347 | **165** | 2 | 353 | **161** | 20.2 | 351 | **159** | 143.3 | 4 | 2% | 6 | 4% | 2 | 1% |
| **CSCB14** | 50 | 1000 | 431 | **137** | 2.1 | 438 | **125** | 17.6 | 436 | **126** | 117.8 | 12 | 9% | 12 | 9% | 0 | 0% |
| **CSCB15** | 50 | 2000 | 708 | **247** | 3.3 | 703 | **250** | 82.0 | 697 | **246** | 182.2 | -3 | -1% | 1 | 0% | 4 | 2% |
| **CSCB16** | 100 | 2000 | 846 | **208** | 3.6 | 809 | **185** | 76.1 | 811 | **182** | 215.7 | 23 | 11% | 26 | 13% | 3 | 2% |
| **Average** | | | | | | | | | | | | 5.7 | 5% | 7.3 | 7% | 1.6 | 2% |

**Note**: [1]NT: Number of trips; [2]NB: number of buses; [3]RT: running time (seconds); [4]PMCM2Park: bus saving of PMCM w.r.t Park; [5]NB: $NB_{Park} - NB_{PMCM}$; [6]%: $PMCM2Park / NB_{Park} \times 100\%$; [7]PMCMPI2Park: bus saving of PMCMPI w.r.t. Park; [8]PI2PMCM: bus saving of PMCMPI w.r.t. PMCM

Table 8 Computation result of PMCM and PMCMPI for benchmark with MRT=5400

| Scenario | # of schools | # of stops | Park | | | PMCM | | | PMCMPI | | | PMCM2Park[4] | | PMCMPI2Park[7] | | PI2PMCM[8] | |
|---|---|---|---|---|---|---|---|---|---|---|---|---|---|---|---|---|---|
| | | | NT[1] | NB[2] | RT[3] | NT | NB | RT | NT | NB | RT | NB[5] | %[6] | NB | % | NB | % |
| **RSRB01** | 6 | 250 | 65 | **31** | 0.5 | 55 | **31** | 0.7 | 56 | **29** | 21.3 | 0 | 0% | 2 | 6% | 2 | 6% |
| **RSRB02** | 12 | 250 | 72 | **30** | 0.5 | 69 | **29** | 1.0 | 69 | **26** | 18.9 | 1 | 3% | 4 | 13% | 3 | 10% |
| **RSRB03** | 12 | 500 | 129 | **61** | 0.8 | 120 | **50** | 2.3 | 118 | **46** | 37.9 | 11 | 18% | 15 | 25% | 4 | 8% |
| **RSRB04** | 25 | 500 | 149 | **56** | 0.9 | 132 | **50** | 3.2 | 131 | **48** | 33.9 | 6 | 11% | 8 | 14% | 2 | 4% |
| **RSRB05** | 25 | 1000 | 268 | **106** | 1.5 | 232 | **88** | 7.6 | 232 | **85** | 122.8 | 18 | 17% | 21 | 20% | 3 | 3% |
| **RSRB06** | 50 | 1000 | 275 | **82** | 1.6 | 225 | **81** | 9.9 | 226 | **78** | 126.8 | 1 | 1% | 4 | 5% | 3 | 4% |
| **RSRB07** | 50 | 2000 | 554 | **159** | 2.4 | 456 | **154** | 35.7 | 457 | **143** | 276.0 | 5 | 3% | 16 | 10% | 11 | 7% |
| **RSRB08** | 100 | 2000 | 714 | **158** | 2.9 | 588 | **141** | 50.2 | 587 | **133** | 281.6 | 17 | 11% | 25 | 16% | 8 | 6% |
| **CSCB01** | 6 | 250 | 72 | **35** | 0.5 | 68 | **34** | 1.2 | 69 | **33** | 19.0 | 1 | 3% | 2 | 6% | 1 | 3% |
| **CSCB02** | 12 | 250 | 66 | **27** | 0.5 | 59 | **25** | 0.9 | 60 | **25** | 16.0 | 2 | 7% | 2 | 7% | 0 | 0% |
| **CSCB03** | 12 | 500 | 133 | **52** | 0.7 | 126 | **50** | 2.1 | 124 | **48** | 39.4 | 2 | 4% | 4 | 8% | 2 | 4% |
| **CSCB04** | 25 | 500 | 151 | **57** | 0.9 | 140 | **48** | 5.2 | 145 | **52** | 37.1 | 9 | 16% | 9 | 16% | 0 | 0% |
| **CSCB05** | 25 | 1000 | 341 | **115** | 1.6 | 299 | **119** | 9.0 | 301 | **113** | 93.7 | -4 | -3% | 2 | 2% | 6 | 5% |
| **CSCB06** | 50 | 1000 | 384 | **123** | 1.9 | 330 | **113** | 7.8 | 331 | **110** | 83.6 | 10 | 8% | 13 | 11% | 3 | 3% |
| **CSCB07** | 50 | 2000 | 560 | **164** | 2.6 | 498 | **152** | 33.3 | 495 | **146** | 312.0 | 12 | 7% | 18 | 11% | 6 | 4% |
| **CSCB08** | 100 | 2000 | 619 | **156** | 2.9 | 560 | **142** | 40.4 | 557 | **138** | 238.7 | 14 | 9% | 18 | 12% | 4 | 3% |
| **CSCB09** | 6 | 250 | 78 | **26** | 0.4 | 69 | **25** | 0.8 | 70 | **21** | 21.4 | 1 | 4% | 5 | 19% | 4 | 16% |
| **CSCB10** | 12 | 250 | 86 | **30** | 0.5 | 78 | **27** | 1.2 | 77 | **24** | 16.9 | 3 | 10% | 6 | 20% | 3 | 11% |
| **CSCB11** | 12 | 500 | 118 | **50** | 0.7 | 104 | **44** | 2.4 | 105 | **43** | 56.4 | 6 | 12% | 7 | 14% | 1 | 2% |
| **CSCB12** | 25 | 500 | 133 | **55** | 0.9 | 120 | **52** | 4.0 | 119 | **52** | 48.2 | 3 | 5% | 3 | 5% | 0 | 0% |
| **CSCB13** | 25 | 1000 | 284 | **135** | 1.7 | 255 | **130** | 11.5 | 254 | **124** | 166.3 | 5 | 4% | 11 | 8% | 6 | 5% |
| **CSCB14** | 50 | 1000 | 329 | **117** | 1.9 | 306 | **98** | 17.4 | 306 | **101** | 80.8 | 19 | 16% | 19 | 16% | 0 | 0% |
| **CSCB15** | 50 | 2000 | 577 | **215** | 2.8 | 486 | **195** | 54.2 | 484 | **191** | 358.7 | 20 | 9% | 24 | 11% | 4 | 2% |
| **CSCB16** | 100 | 2000 | 582 | **158** | 2.9 | 489 | **158** | 52.5 | 493 | **152** | 206.8 | 0 | 0% | 6 | 4% | 6 | 4% |
| **Average** | | | | | | | | | | | | 6.8 | 7% | 10.2 | 12% | 3.4 | 5% |

**Note**: [1]NT: Number of trips; [2]NB: number of buses; [3]RT: running time (seconds); [4]PMCM2Park: bus saving of PMCM w.r.t Park; [5]NB: $NB_{Park} - NB_{PMCM}$; [6]%: $PMCM2Park / NB_{Park} \times 100\%$; [7]PMCMPI2Park: bus saving of PMCMPI w.r.t. Park; [8]PI2PMCM: bus saving of PMCMPI w.r.t. PMC

In loose MRT scenarios (MRT=5400), the bus savings are significant (Table 8 and Figure 6). The highest bus saving for PMCM occurred at CSCB15 where 20 buses can be saved, which equals 9% of total buses. The highest bus saving ratio can reach 18% just due to PMCM. On average, PMCM can saves 6.8 buses compared to Park, which is a 7% improvement. The results of PMCMPI were even better. The average bus saving of PMCMPI was 12%, which is equivalent to 10.2 bus saving. The highest bus saving reached 25% at RSRB03 where 15 buses can be reduced out of 61 buses. At larger case like RSRB08, 16% improvement saves 25 buses. Thanks to the larger feasible region resulted by loosening the MRT constraints, PI played a more critical role in routing than in stricter MRT condition. On average, PI can reduce buses by 5% (which equals to about 3.4 buses) from PMCM. The significant performance of the MCMPI over the benchmark solutions comes from the 'look ahead' idea of incorporating the scheduling information into the routing problem. With the consideration of trip compatibility, the routing plan is more likely to provide a scheduling plan than the traditional methods.

## 5. Conclusion

In this paper, an insertion based Minimum Cost Matching with Post Improvement (MCMPI) algorithm was proposed to solve multi-school bus routing problems. It adopts the idea of incorporating trip compatibility into the routing problem. Different experiments were conducted to compare different variations of the algorithm. We showed that the heuristic algorithm can find solutions with higher quality than the exact methods in much shorter time. In large size problems, where the exact method cannot find good solutions, the heuristic algorithm still performs efficiently. Other experiments were conducted based on the benchmark problems from Park et al. (2012). The results show that the proposed Minimum Cost Matching (MCM) heuristic algorithm could save up to 25% of buses compared to Park et al. (2012). On average, the newly proposed algorithms can save 7% and 12% of buses at strict Maximum Ride Time (MRT=2700 seconds) conditions and loose MRT (=5400 seconds) cases, respectively. Moreover, the post-improvement (PI) can further improve the solution from MCM by up to 16%.

This study opens avenues for future research on school bus routing. One of them is to incorporate sources of unreliability such as travel time uncertainties in the routing process. Another one is finding the optimal initial number of trips required for the MCM. Another significant step could be incorporating the school bell/dismissal time problem in routing and solving these two problems together since the school bell times affect the trip compatibilities.